
\def\prd#1#2#3#4{#1, {\sl Phys. Rev. \/}{\bf D#2}, #3 (19#4)}

\def\pr#1#2#3#4{#1, {\sl Phys. Rev. \/}{\bf #2}, #3 (19#4)}

\def\aar#1#2#3#4{#1, {\sl Astron. Astrophys. Rev. \/}{\bf #2}, #3 (19#4)}

\def\npb#1#2#3#4{#1, {\sl Nucl. Phys. \/}{\bf B#2}, #3 (19#4)}

\def\apj#1#2#3#4{#1, {\sl Astrophys. J. \/} {\bf#2}, #3 (19#4)}
\def\apj#1#2#3#4{#1, {\sl Astrophys. J. \/} {\bf#2}, #3 (19#4)}

\def\grg#1#2#3#4{#1, {\sl Gen. Rel. Grav. \/}{\bf #2}, #3 (19#4)}

\def\aar#1#2#3#4{#1, {\sl Astron. Astrophys. Rev. \/}{\bf #2}, #3 (19#4)}

\def\obscure#1#2#3#4#5{#1, {\sl #2\/} {\bf #3}, #4 (19#5)}
\def\inbooked#1#2#3#4#5{#1, in {\it #2\/,} ed. #3 ({#4,} 19#5)}

\def\book#1#2#3#4{#1, {\it #2\/} (#3, 19#4)}

\font\twelverm=cmr10 scaled 1200    \font\twelvei=cmmi10 scaled 1200
\font\twelvesy=cmsy10 scaled 1200   \font\twelveex=cmex10 scaled 1200
\font\twelvebf=cmbx10 scaled 1200   \font\twelvesl=cmsl10 scaled 1200
\font\twelvett=cmtt10 scaled 1200   \font\twelveit=cmti10 scaled 1200

\skewchar\twelvei='177   \skewchar\twelvesy='60


\def\twelvepoint{\normalbaselineskip=12.4pt
  \abovedisplayskip 12.4pt plus 3pt minus 9pt

  \belowdisplayskip 12.4pt plus 3pt minus 9pt
  \abovedisplayshortskip 0pt plus 3pt
  \belowdisplayshortskip 7.2pt plus 3pt minus 4pt
  \smallskipamount=3.6pt plus1.2pt minus1.2pt
  \medskipamount=7.2pt plus2.4pt minus2.4pt
  \bigskipamount=14.4pt plus4.8pt minus4.8pt
  \def\rm{\fam0\twelverm}          \def\it{\fam\itfam\twelveit}%
  \def\sl{\fam\slfam\twelvesl}     \def\bf{\fam\bffam\twelvebf}%
  \def\mit{\fam 1}                 \def\cal{\fam 2}%
  \def\tt{\twelvett}
  \def\nullspace{\nulldelimiterspace=0pt \mathsurround=0pt }
  \def\big##1{{\hbox{$\left##1\vbox to 10.2pt{}\right.\nullspace$}}}
  \def\Big##1{{\hbox{$\left##1\vbox to 13.8pt{}\right.\nullspace$}}}
  \def\bigg##1{{\hbox{$\left##1\vbox to 17.4pt{}\right.\nullspace$}}}
  \def\Bigg##1{{\hbox{$\left##1\vbox to 21.0pt{}\right.\nullspace$}}}
  \textfont0=\twelverm   \scriptfont0=\tenrm   \scriptscriptfont0=\sevenrm
  \textfont1=\twelvei    \scriptfont1=\teni    \scriptscriptfont1=\seveni
  \textfont2=\twelvesy   \scriptfont2=\tensy   \scriptscriptfont2=\sevensy
  \textfont3=\twelveex   \scriptfont3=\twelveex  \scriptscriptfont3=\twelveex
  \textfont\itfam=\twelveit
  \textfont\slfam=\twelvesl
  \textfont\bffam=\twelvebf \scriptfont\bffam=\tenbf

  \scriptscriptfont\bffam=\sevenbf
  \normalbaselines\rm}



\def\beginlinemode{\endmode
  \begingroup\parskip=0pt \obeylines\def\\{\par}\def\endmode{\par\endgroup}}
\def\beginparmode{\endmode
  \begingroup \def\endmode{\par\endgroup}}
\let\endmode=\par
{\obeylines\gdef\
{}}
\def\singlespace{\baselineskip=\normalbaselineskip}
\def\oneandathirdspace{\baselineskip=\normalbaselineskip
  \multiply\baselineskip by 4 \divide\baselineskip by 3}
\def\oneandahalfspace{\baselineskip=\normalbaselineskip
  \multiply\baselineskip by 3 \divide\baselineskip by 2}
\def\doublespace{\baselineskip=\normalbaselineskip \multiply\baselineskip by 2}

\newcount\firstpageno
\firstpageno=2
\footline={\ifnum\pageno<\firstpageno{\hfil}\else{\hfil\twelverm\folio\hfil}\fi}
\let\rawfootnote=\footnote
\def\footnote#1#2{{\tenrm\singlespace\parindent=0pt\rawfootnote{#1}{#2}}}
\def\raggedcenter{\leftskip=4em plus 12em \rightskip=\leftskip
  \parindent=0pt \parfillskip=0pt \spaceskip=.3333em \xspaceskip=.5em
  \pretolerance=9999 \tolerance=9999
  \hyphenpenalty=9999 \exhyphenpenalty=9999 }
\def\dateline{\rightline{\ifcase\month\or
  January\or February\or March\or April\or May\or June\or
  July\or August\or September\or October\or November\or December\fi
  \space\number\year}}
\def\received{\vskip 3pt plus 0.2fill
 \centerline{\sl (Received\space\ifcase\month\or
  January\or February\or March\or April\or May\or June\or
  July\or August\or September\or October\or November\or December\fi
  \qquad, \number\year)}}


\hsize=6.5truein
\vsize=8.9truein
\parskip=\smallskipamount
\twelvepoint
\doublespace
\overfullrule=0pt


\def\preprintno#1{\singlespace
 \rightline{\rm #1}}

\def\title
  {\null\vskip 3pt plus 0.2fill
   \beginlinemode \doublespace \raggedcenter \bf}

\font\titlefont=cmr10 scaled \magstep3
\def\bigtitle{\null\vskip 3pt plus 0.2fill \beginlinemode \doublespace
\raggedcenter \titlefont}


\font\twelvesc=cmcsc10 scaled 1200
\def\author{\vskip 16pt plus 0.2fill \beginlinemode\singlespace
\raggedcenter\twelvesc}

\def\affil
  {\vskip 4pt plus 0.1fill \beginlinemode
   \oneandahalfspace \raggedcenter \sl}

\def\abstract  
  {\vskip 24pt plus 0.3fill \beginparmode
   \narrower\centerline{ABSTRACT}\vskip 12pt }

\def\endtitlepage{\endpage\body}

\def\body{\beginparmode}

\def\head#1{
  \filbreak\vskip 0.35truein
  {\immediate\write16{#1}
   \raggedcenter \uppercase{#1}\par}
   \nobreak\vskip 0.2truein\nobreak}

\def\subhead#1{
  \vskip 0.20truein
  {\raggedcenter #1 \par}
   \nobreak\vskip 0.15truein\nobreak}

\def\References  
  {\subhead{References}
   \beginparmode
   \frenchspacing\parindent=0pt \leftskip=1truecm
   \parskip=2pt plus 3pt \everypar{\hangindent=\parindent}}

\def\publications   
  {\beginparmode
   \frenchspacing \parindent=0pt \leftskip=1truecm
   \parskip=2pt plus 3pt \everypar{\hangindent=\parindent}}

\gdef\r#1{\indent\hbox to 0pt{\hss#1.~}}   

\gdef\refis#1{\indent\hbox to 0pt{\hss[#1]~}}   

\def\endreferences{\body}

\def\figurecaptions
  {\endpage
   \beginparmode
   \head{Figure Captions}
}

\def\endpage
  {\vfill\eject}

\def\endpaper
  {\endmode\vfill\supereject}

\def\endit
  {\endpaper\end}


\def\cite#1{{#1}}
\def\[#1]{[\cite{#1}]}                  
\def\q#1{\ \[#1]}			
\def\refto#1{$^{#1}$}                   
\def\ref#1{Ref.~#1}                     
\def\Ref#1{Ref.~#1}                     

\def\call#1{{#1}}
\def\(#1){(\call{#1})}

\def\taghead#1{}
\def\frac#1#2{{#1 \over #2}}
\def\half{{\frac 12}}

\def\12{{1\over 2}}
\def\eg{{\it e.g.,\ }}

\def\ie{{\it i.e.,\ }}

\def\etc{{\it etc.\ }}

\def\sla{\raise.15ex\hbox{$/$}\kern-.57em}

\def\leaderfill{\leaders\hbox to 1em{\hss.\hss}\hfill}
\def\twiddle{\lower.9ex\rlap{$\kern-.1em\scriptstyle\sim$}}
\def\bigtwiddle{\lower1.ex\rlap{$\sim$}}
\def\gtwid{\mathrel{\raise.3ex\hbox{$>$\kern-.75em\lower1ex\hbox{$\sim$}}}}
\def\ltwid{\mathrel{\raise.3ex\hbox{$<$\kern-.75em\lower1ex\hbox{$\sim$}}}}
\def\square{\kern1pt\vbox{\hrule height 1.2pt\hbox{\vrule width 1.2pt\hskip 3pt
   \vbox{\vskip 6pt}\hskip 3pt\vrule width 0.6pt}\hrule height 0.6pt}\kern1pt}
\def\ucsb{Department of Physics\\University of California\\
Santa Barbara, CA 93106}
\def\hu{Racah Institute of Physics\\Hebrew University of Jerusalem\\
Givat Ram, Jerusalem 91904\\Israel}

\def\m@th{\mathsurround=0pt }
\def\leftrightarrowfill{$\m@th \mathord\leftarrow \mkern-6mu
 \cleaders\hbox{$\mkern-2mu \mathord- \mkern-2mu$}\hfill
 \mkern-6mu \matide\baselineskip by 5}

\def\martinstyletitle                      
  {\null\vskip 3pt plus 0.2fill
   \beginlinemode \doublespace \raggedcenter \titlefont}

\catcode`@=11
\newcount\tagnumber\tagnumber=0

\immediate\newwrite\eqnfile
\newif\if@qnfile\@qnfilefalse
\def\write@qn#1{}
\def\writenew@qn#1{}
\def\w@rnwrite#1{\write@qn{#1}\message{#1}}
\def\@rrwrite#1{\write@qn{#1}\errmessage{#1}}

\def\taghead#1{\gdef\t@ghead{#1}\global\tagnumber=0}
\def\t@ghead{}

\expandafter\def\csname @qnnum-3\endcsname
  {{\t@ghead\advance\tagnumber by -3\relax\number\tagnumber}}
\expandafter\def\csname @qnnum-2\endcsname
  {{\t@ghead\advance\tagnumber by -2\relax\number\tagnumber}}
\expandafter\def\csname @qnnum-1\endcsname
  {{\t@ghead\advance\tagnumber by -1\relax\number\tagnumber}}
\expandafter\def\csname @qnnum0\endcsname
  {\t@ghead\number\tagnumber}
\expandafter\def\csname @qnnum+1\endcsname
  {{\t@ghead\advance\tagnumber by 1\relax\number\tagnumber}}
\expandafter\def\csname @qnnum+2\endcsname
  {{\t@ghead\advance\tagnumber by 2\relax\number\tagnumber}}
\expandafter\def\csname @qnnum+3\endcsname
  {{\t@ghead\advance\tagnumber by 3\relax\number\tagnumber}}

\def\equationfile{%
  \@qnfiletrue\immediate\openout\eqnfile=\jobname.eqn%
  \def\write@qn##1{\if@qnfile\immediate\write\eqnfile{##1}\fi}
  \def\writenew@qn##1{\if@qnfile\immediate\write\eqnfile
    {\noexpand\tag{##1} = (\t@ghead\number\tagnumber)}\fi}
}

\def\callall#1{\xdef#1##1{#1{\noexpand\call{##1}}}}
\def\call#1{\each@rg\callr@nge{#1}}

\def\each@rg#1#2{{\let\thecsname=#1\expandafter\first@rg#2,\end,}}
\def\first@rg#1,{\thecsname{#1}\apply@rg}
\def\apply@rg#1,{\ifx\end#1\let\next=\relax%
\else,\thecsname{#1}\let\next=\apply@rg\fi\next}

\def\callr@nge#1{\calldor@nge#1-\end-}
\def\callr@ngeat#1\end-{#1}
\def\calldor@nge#1-#2-{\ifx\end#2\@qneatspace#1 %
  \else\calll@@p{#1}{#2}\callr@ngeat\fi}
\def\calll@@p#1#2{\ifnum#1>#2{\@rrwrite{Equation range #1-#2\space is bad.}
\errhelp{If you call a series of equations by the notation M-N, then M and
N must be integers, and N must be greater than or equal to M.}}\else%
 {\count0=#1\count1=#2\advance\count1
by1\relax\expandafter\@qncall\the\count0,%
  \loop\advance\count0 by1\relax%
    \ifnum\count0<\count1,\expandafter\@qncall\the\count0,%
  \repeat}\fi}

\def\@qneatspace#1#2 {\@qncall#1#2,}
\def\@qncall#1,{\ifunc@lled{#1}{\def\next{#1}\ifx\next\empty\else
  \w@rnwrite{Equation number \noexpand\(>>#1<<) has not been defined yet.}
  >>#1<<\fi}\else\csname @qnnum#1\endcsname\fi}

\let\eqnono=\eqno
\def\eqno(#1){\tag#1}
\def\tag#1$${\eqnono(\displayt@g#1 )$$}

\def\aligntag#1\endaligntag
  $${\gdef\tag##1\\{&(##1 )\cr}\eqalignno{#1\\}$$
  \gdef\tag##1$${\eqnono(\displayt@g##1 )$$}}

\def\eqalignno#1{\displ@y \tabskip\centering
  \halign to\displaywidth{\hfil$\displaystyle{##}$\tabskip\z@skip
    &$\displaystyle{{}##}$\hfil\tabskip\centering
    &\llap{$\displayt@gpar##$}\tabskip\z@skip\crcr
    #1\crcr}}

\def\displayt@gpar(#1){(\displayt@g#1 )}

\def\displayt@g#1 {\rm\ifunc@lled{#1}\global\advance\tagnumber by1
        {\def\next{#1}\ifx\next\empty\else\expandafter
        \xdef\csname @qnnum#1\endcsname{\t@ghead\number\tagnumber}\fi}%
  \writenew@qn{#1}\t@ghead\number\tagnumber\else
        {\edef\next{\t@ghead\number\tagnumber}%
        \expandafter\ifx\csname @qnnum#1\endcsname\next\else
        \w@rnwrite{Equation \noexpand\tag{#1} is a duplicate number.}\fi}%
  \csname @qnnum#1\endcsname\fi}

\def\ifunc@lled#1{\expandafter\ifx\csname @qnnum#1\endcsname\relax}

\let\@qnend=\end\gdef\end{\if@qnfile
\immediate\write16{Equation numbers written on []\jobname.EQN.}\fi\@qnend}

\catcode`@=12


\catcode`@=11
\newcount\r@fcount \r@fcount=0
\newcount\r@fcurr
\immediate\newwrite\reffile
\newif\ifr@ffile\r@ffilefalse
\def\w@rnwrite#1{\ifr@ffile\immediate\write\reffile{#1}\fi\message{#1}}

\def\writer@f#1>>{}
\def\referencefile{
  \r@ffiletrue\immediate\openout\reffile=\jobname.ref%
  \def\writer@f##1>>{\ifr@ffile\immediate\write\reffile%
    {\noexpand\refis{##1} = \csname r@fnum##1\endcsname = %
     \expandafter\expandafter\expandafter\strip@t\expandafter%
     \meaning\csname r@ftext\csname r@fnum##1\endcsname\endcsname}\fi}%
  \def\strip@t##1>>{}}

\def\citeall#1{\xdef#1##1{#1{\noexpand\cite{##1}}}}
\def\cite#1{\each@rg\citer@nge{#1}}

\def\each@rg#1#2{{\let\thecsname=#1\expandafter\first@rg#2,\end,}}
\def\first@rg#1,{\thecsname{#1}\apply@rg}
\def\apply@rg#1,{\ifx\end#1\let\next=\relax
\else,\thecsname{#1}\let\next=\apply@rg\fi\next}

\def\citer@nge#1{\citedor@nge#1-\end-}
\def\citer@ngeat#1\end-{#1}
\def\citedor@nge#1-#2-{\ifx\end#2\r@featspace#1 
  \else\citel@@p{#1}{#2}\citer@ngeat\fi}
\def\citel@@p#1#2{\ifnum#1>#2{\errmessage{Reference range #1-#2\space is bad.}%
    \errhelp{If you cite a series of references by the notation M-N, then M and
    N must be integers, and N must be greater than or equal to M.}}\else%
 {\count0=#1\count1=#2\advance\count1
by1\relax\expandafter\r@fcite\the\count0,%
  \loop\advance\count0 by1\relax
    \ifnum\count0<\count1,\expandafter\r@fcite\the\count0,%
  \repeat}\fi}

\def\r@featspace#1#2 {\r@fcite#1#2,}
\def\r@fcite#1,{\ifuncit@d{#1}
    \newr@f{#1}%
    \expandafter\gdef\csname r@ftext\number\r@fcount\endcsname%
                     {\message{Reference #1 to be supplied.}%
                      \writer@f#1>>#1 to be supplied.\par}%
 \fi%
 \csname r@fnum#1\endcsname}
\def\ifuncit@d#1{\expandafter\ifx\csname r@fnum#1\endcsname\relax}%
\def\newr@f#1{\global\advance\r@fcount by1%
    \expandafter\xdef\csname r@fnum#1\endcsname{\number\r@fcount}}

\let\r@fis=\refis
\def\refis#1#2#3\par{\ifuncit@d{#1}
   \newr@f{#1}%
   \w@rnwrite{Reference #1=\number\r@fcount\space is not cited up to now.}\fi%
  \expandafter\gdef\csname r@ftext\csname r@fnum#1\endcsname\endcsname%
  {\writer@f#1>>#2#3\par}}

\def\ignoreuncited{
   \def\refis##1##2##3\par{\ifuncit@d{##1}%
     \else\expandafter\gdef\csname r@ftext\csname
r@fnum##1\endcsname\endcsname%
     {\writer@f##1>>##2##3\par}\fi}}

\def\r@ferr{\endreferences\errmessage{I was expecting to see
\noexpand\endreferences before now;  I have inserted it here.}}
\let\r@ferences=\references
\def\references{\r@ferences\def\endmode{\r@ferr\par\endgroup}}

\let\endr@ferences=\endreferences
\def\endreferences{\r@fcurr=0
  {\loop\ifnum\r@fcurr<\r@fcount
    \advance\r@fcurr by 1\relax\expandafter\r@fis\expandafter{\number\r@fcurr}%
    \csname r@ftext\number\r@fcurr\endcsname%
  \repeat}\gdef\r@ferr{}\endr@ferences}


\let\r@fend=\endpaper\gdef\endpaper{\ifr@ffile
\immediate\write16{Cross References written on []\jobname.REF.}\fi\r@fend}

\catcode`@=12

\citeall\refto
\citeall\ref%
\citeall\Ref%


\def\i#1{{\it #1\/}} 
\def\spose#1{\hbox to 0pt{#1\hss}}
\def\-{\ifmmode \widetilde \else \~\fi}

\def\const{{\rm const.\,}}
\def\or{{\rm\ \ or\ \ }}

\def\and{{\rm\ \ and\ \ }}





\def\rarrow{\rightarrow}

\def\Dt{\spose{\raise 1.5ex\hbox{\hskip3pt$\mathchar"201$}}}   

\def\dt{\spose{\raise 1.0ex\hbox{\hskip2pt$\mathchar"201$}}}    

\def\lta{\mathrel{\s pose{\lower 3pt\hbox{$\mathchar"218$}}\raise
2.0pt\hbox{$\mathchar"13C$}}}
\def\gta{\mathrel{\s pose{\lower 3pt\hbox{$\mathchar"218$}}
\raise 2.0pt\hbox{$\mathchar"13E$}}}

\def\simlt{\lower.5ex\hbox{\ltsima}}

\def\simgt{\lower.5ex\hbox{\gtsima}}


\def\al{_\alpha{}}
\def\be{_\beta{}}
\def\ga{_\gamma{}}
\def\de{_\delta{}}

\def\m{_\mu{}}
\def\n{_\nu{}}

\def\Al{^\alpha{}}
\def\Be{^\beta{}}
\def\Ga{^\gamma{}}
\def\De{^\delta{}}

\def\M{^\mu{}}
\def\N{^\nu{}}


\def\gab{g_{\alpha\beta}}
\def\gmn{g_{\mu\nu}}
\def\gAB{g^{\alpha\beta}}

\def\gtab{\-g_{\alpha\beta}}

\def\gtAB{\-g^{\alpha\beta}}

\def\lineab{\gab\,dx\Al dx\Be}
\def\linemn{\gmn\,dx\M dx\N}
\def\plineab{\gtab\,dx\Al dx\Be}

\def\G{{\cal G}}
\def\bigtitle{\null\vskip 3pt plus 0.2fill \beginlinemode \doublespace
\raggedcenter \titlefont}
\singlespace
\preprintno{UCSB-TH-92-41}
\preprintno{November 1992}
\doublespace
\bigtitle The Relation between Physical and
Gravitational Geometry
\author
Jacob D. Bekenstein\footnote{$^1$}{E-mail: jacob@cosmic.physics.ucsb.edu\hfil}
\affil\ucsb
\medskip
{\rm  and}
\medskip
\hu
\centerline{\hfil UCSB-TH-92-41}

\abstract
\oneandathirdspace\parskip=2pt plus 3pt
The appearance of two geometries in a
single gravitational theory is familiar.            Usually, as in the
Brans-Dicke theory or in string theory, these are conformally related
Riemannian geometries.  Is this the most general relation between the
two geometries allowed by physics ?   We study   this question by
supposing  that  the physical geometry on which matter dynamics take place
could be Finslerian rather than just Riemannian.
An appeal to the weak equivalence principle and causality then leads us
to the conclusion that
the Finsler geometry has to reduce to   a Riemann geometry whose metric - the
physical metric -  is related to the gravitational metric by a generalization
of the conformal
transformation.
\oneandathirdspace
\parskip=2pt plus 3pt
\subhead{{\bf I. INTRODUCTION}}

The excellent description provided by special relativity of elementary
particle phenomena is usually taken to imply that spacetime is described
by a Riemannian geometry.  This is because special relativity implies a
Minkowski geometry for spacetime,
but                as shown by Schild\q{Schild}, the experimental
existence of the gravitational redshift makes it impossible for
the Minkowski geometry to apply globally.  An obvious way to mesh the
Minkowski geometries at
various points is to have a global Riemannian geometry which the
Minkowski geometry of elementary particle physics is tangent to at each
spacetime event.
This is the situation in a one-geometry description of physics,
 \eg
general relativity (GR).

However, physics may not be that simple: gravitation may naturally require two
geometries for its descrption.
Two geometries in a single  theory made their debut in Nordstr{\o}m's 1913
gravitational theory\q{Nordstrom} which preceded
GR.  As in Nordstr{\o}m's theory, so in theories like Brans-Dicke\q{Dicke}, the
variable mass theory\q{VMT},    Dirac's theory of a variable gravitational
constant\q{Dirac}, string theories\q{strings} and many others,
 two conformally
related geometries appear.  Usually one of these describes
gravitation while the other defines the geometry in which matter
plays out its dynamics.  The strong equivalence principle
is violated by all these two-geometries theories, but they usually preserve
weak equivalence.
Theories of these sort have been of great value in
clarifying the foundations of gravitation theory.

Thus, the two-geometries approach to the formulation of gravitational theory is
an important paradigm.  Whenever it becomes necessary to formulate a new
theory of gravity,  a conservative way to proceed  in order to avoid
immediate conflict with the tests of GR  is to invoke a Riemannian metric
$\gab$, build the
Einstein-Hilbert action  for  the geometry's dynamics out of it, and effect the
departure from standard GR by  prescribing the relation between  $\gab$ and the
physical
geometry on which matter propagates.    Most known theories assume the relation
is  a
simple conformal transformation.

However, the conformal transformation is but the simplest way to relate two
geometries.  Might the relation between gravitational and physical
geometries be more complicated ?  In other words, within the two-geometries
paradigm
for gravitational theory, what are the most general theories that may be
envisaged ?  To
answer this question we consider physical geometry of the most general kind
that might be
of interest physically, namely Finslerian geometry rather than Riemannian one.

Finsler geometry, introduced in Finsler's 1918 G\"ottingen dissertation, is the
most general
geometry in which the squared line element is  homogeneous of second degree  in
the
coordinate increments\q{Cartan}:
$$
ds^2=f(x\Al, dx\Be); \qquad     f(x\Al, \mu dx\Be)=\mu^2 f(x\Al, dx\Be).
\eqno(finsler)
$$
Whenever $f(x\Al, dx\Be)$ is a quadratic form in the $dx\Be$, the geometry is
Riemannian.  Otherwise, we have a (rather complicated) Finsler geometry as the
playground of matter dynamics. We shall show presently that this state of
affairs cannot be
ruled out at the outset by the argument about the Minkowski geometry of
elementary
particle physics.

Although Finsler geometry is quite different from Riemann geometry, it
is possible to introduce a metric-like tensor for it.  If in the second of
Eqs.~\(finsler) we
replace $\mu\rarrow 1+\epsilon$ \ where $\epsilon$ is an infinitesimal, expand
in
$\epsilon$, and focus on the terms of ${\rm O}(\epsilon^2)$, we have (this is a
facet of
Euler's theorem)
$$
\G\al\be\, dx\Al dx\Be  \equiv\half {\partial^2 f\over \partial dx\Al
 \partial dx\Be}\, dx\Al dx\Be   = f .
\eqno(Euler)
$$
Thus we may express the line element as\q{Roxburgh}
$$
ds^2 = \G\al\be\,  dx\Al dx\Be ,   \eqno(quasimetric)
$$
and think of ${\cal G}\al\be$ as a kind of metric.  However, it must be
remembered that $\G\al\be$ itself depends on $dx\Al$. Because of this
difference from the Riemann metric, we call $\G\al\be$ the quasimetric.
Our arguments will make heavy use of it.

\subhead{{\bf II. THE SPIRIT OF  COVARIANCE}}

Because in the theory being discussed there is already a
symmetric tensor, $\gab$, we
may rewrite
Eq.~\(finsler) in general    as follows:
$$
ds^2=\gab\, dx\Al dx\Be\, {\cal F}(x\Al, dx^1/dx^0, dx^2/dx^0, dx^3/dx^0).
\eqno(simplefinsler)
$$
This is because the expressions like $dx^1/dx^0$ are the only independent
combinations of the coordinate increments which are homogeneous of
degree zero in $dx\Al$.

As it stands, Eq.~\(simplefinsler) is still the most general Finsler
geometry.  However, there is something about it which does violence to
the spirit of the principle of covariance.  Suppose we make a general
coordinate transformation.  We know that the form $\gab\, dx\Al dx\Be$ is
invariant.  This is achieved by the components of $\gab$ changing in an
appropriate way so that at a fixed spacetime point the metric is
described by ten numbers of which four may be chosen arbitrarily.  In
other words, covariance requires                 that four numbers be
free at any spacetime point.             Since we also want
invariance of $ds^2$, not just of $\gab\, dx\Al dx\Be$, it is plain that
the form of the function ${\cal F}$ cannot be invariant.  It will vary with
coordinate systems in such a way as to
compensate for the transformations of the ratios $dx^i/dx^0$ into
rational functions of themselves.

Not only is this ugly, but it also means that the freedom inherent in
coordinate transformations is, at a fixed point in spacetime, not just
that in
four numbers, but rather that in a function of three variables.
Although the letter of the principle of covariance is still obeyed by
having this free function ${\cal F}$, it would seem that the spirit of the
principle is violated.  The vast freedom engendered by
coordinate transformations would seem to empty the principle of any physical
content.

One may recover the situation where only a few quantities are free at
a point by confining attention to a function $F$ of coordinate invariants
alone.
However, out of the invariant $\gab\,dx\Al dx\Be$  alone one can form
only one homogeneous function of second order in the $dx\Al$: the invariant
itself. To go beyond triviality one needs more invariants.  Now if there exists
a dimensionless scalar field $\psi$ (like the dilaton in many
contemporary field theories) and a length scale $L$,  \eg  the Planck length, a
nontrivial
Finsler line element may be written:
$$
ds^2=\gab\, dx\Al dx\Be\,F(I, H, \psi) ,   \eqno(condensedfinsler)
$$
$$
I\equiv L^2\gAB\psi,\al \psi,\be ,     \eqno(I)
$$
$$
H\equiv L^2\,{(\psi,\al dx\Al)^2\over -\gab\, dx\Al dx\Be}.    \eqno(H)
$$
Note that both $F$ and its arguments $I$, $H$ and $\psi$ are
dimensionless.

Of course, we could have added other arguments to $F$
constructed out of second derivatives of $\psi$.  We refrain from this
in order to preclude higher derivative terms from entering in the matter
equations of motion (after all the matter action will be built on the
line element $ds^2$).
The introduction of more scalar fields as arguments of $F$ is not
logically excluded.  However, given that $\psi$ is a field with a
special status (building block of the physical geometry), simplicity
requires that we abstain from introducing more such entities.
With this proviso our line element is the most
general that may be constructed solely out of coordinate invariants.

We now note that covariance of  Eq.~\(condensedfinsler) is to be had at the
same price as that for ordinary Riemannian geometry: four free metric
components at every point in spacetime.  The function $F$ is fixed, one
and the same for all coordinate systems.

It is consistent with all our previous discussion to postulate that the
classical trajectories of free particles are those which extremize the action
$$
S=\half\int{\gab\, \dt x\Al \dt x\Be\,
 F(I, H, \psi)\,d\lambda} ,  \eqno(action)
$$
Here by $H$ we mean the expression in Eq.~\(H) with $dx\Al\rarrow
\dt x\Al\equiv dx\Al/d\lambda$ ($\lambda$ is a parameter along the
trajectory).   Eq.~\(action) is
the straightforward generalization to Finsler geometry of the action used
in GR for classical particles (in a form not invariant under changes of
parameter $\lambda$). It is easy to see that a trajectory with $ds=0$
all along it
automatically extremizes $S$ [which is simply the integral of
$(ds/d\lambda)^2$].  Thus in this
theory, just as in GR, null curves are automatically trajectories of
free particles.  We do not require that the trajectories which extremize $S$ in
the
Finsler geometry coincide with the geodesics of $\gab$.  There is no
physical basis for such an assumption in our context: the metric $\gab$
is for gravitational phenomena, whereas the Finsler geometry is for matter
dynamics.

Before passing on let us rebut the argument that infers a global Riemannian
geometry from the Minkowski geometry of elementary particle physics.
Consider Eq.~\(condensedfinsler).  For general $F$ the line element
certainly does not look like one that could locally be brought to
Minkowski form by a coordinate transformation.  However, suppose
$F(I,H,\psi)$ is regular and nonvanishing in the limit $I\rarrow 0$
and $H\rarrow 0$.  Then in a region where the $\psi$ field varies slowly
(pressumably the solar system is like that), the line element is seen to
be of the form $ds^2\approx F(0,0,\psi)\,\gab\, dx\Al dx\Be$ which
corresponds to a Riemann geometry.  Therefore, under
everyday circumstances some Finslerian geometries can masquerade as a
Riemannian ones, and can thus be consistent with the evidence from
elementary particle physics.

\subhead{{\bf III. THE FORM OF THE FINSLER FUNCTION} $F$}

As long as all we have to deal with is classical particle motion,
Finsler geometry seems perfectly adequate as the physical geometry.
However, classical physics also involves field equations, \ie Maxwell's,
and once we enter into quantum physics even particles must be discussed
in terms of wave (field) equations. But it is unclear how to formulate
the familiar field equations on a general Finsler geometry.  The problem
is that to formulate the typical field equation  (think of the scalar
equation for concreteness), one requires a contravariant ``metric'' to
raise indeces of derivatives and so form divergences.  To put it another
way, to form a scalar action out of derivatives of fields, one requires
an object capable of raising indeces.  In Riemannian geometry $\gAB$
serves this purpose.   If we try to use the inverse of the Finsler
quasimetric, $\G\Al\Be$, we will find in general that it depends on the
increments $dx\Al$ and not just on the spacetime point.  Clearly, field
equations constructed with $\G\Al\Be$ would be meaningless in general.

One way out of this problem is to  confine attention to geometries for
which $\G\al\be$ is independent of $dx\Al$; this guarantees that
$\G\Al\Be$ will be too.  Let us use definition \(Euler) to write the
quasimetric for the Finsler geometry of Eq.~\(condensedfinsler).  The result is
$$
\G\al\be=(F-HF')\gab -L^2\,(F'+2HF'')\psi,\al\psi,\be
$$
$$-2H^2F''\Big[{\psi,_{(\alpha}
g_{\beta)\mu} dx\M\over \psi,\m dx\M}\, -\, {g\al\m dx\M\,
g\be\n dx\N \over  g\m\n dx\M dx\N}\Big] ,  \eqno(Gab)
$$
where a prime denotes a partial derivative with respect to $H$ and round
brackets around subscripts denote symmetrization.
This $\G\al\be$ will
be independent of $dx\Al$ only if $F''$ vanishes, \ie if
$$
F=A(I,\psi)-B(I,\psi)H, \eqno(linearF)
$$
with $A$ and $B$ two dimensionless functions of the shown arguments.  With an
$F$
like this it is possible to construct the familiar field equations.

When we substitute Eq.~\(linearF) into Eq.~\(condensedfinsler)
we find that the quasimetric reduces to a Riemann metric $\gtab$:
$$
ds^2=\gtab\, dx\Al dx\Be\equiv (\gab A+L^2\,B\,\psi,\al \psi,\be)\,dx\Al dx\Be
{}.
\eqno(geff)
$$
But here the relation between the gravitational metric $\gab$ and the
physical metric $\gtab$ is more complex than via a conformal
transformation.

    Might not  a more general relation between physical and gravitational
geometry be
possible than that we have exhibited ?   Perhaps reasonable matter field
equations can be
formulated on the basis of some structure other than the quasimetric.  For
example,
consider a Finsler geometry determined by a symmetric fourth rank
tensor\q{Roxburgh2}:
$$
ds^4={\cal E}\al\be\ga\de\, dx\Al dx\Be dx\Ga dx\De. \eqno(Rox)
$$
We assume ${\cal E}\al\be\ga\de$ is not degenerate, \ie it cannot be written as
$q\al\be\,q\ga\de$.    Suppose we use the inverse tensor ${\cal E}\Al\Be\Ga\De$
in lieu of a metric to construct invariant field lagrangians.   For a scalar
field $\Phi$ the
simplest choice for the lagrangian,
${\cal L}={\cal E}\Al\Be\Ga\De\,\Phi,\al\Phi,\be\Phi,\ga\Phi,\de$, is quartic
in the
field.  It cannot lead to a linear field equations.  And if we take the square
root of the above
invariant as the lagrangian, linearity is still out of reach.   The full
symmetry of  ${\cal
E}\Al\Be\Ga\De$ also prevents us from forming a lagrangian for an antisymmetric
Maxwell field $F\al\be$ which is quadratic in the field.    A fourth order
invariant can be
built, but even if we use its square root as lagrangian, we cannot obtain
linear equations.
Needless to say, we cannot do  without linear equations in physics.  And in the
more
general case when  $ds^n$ is given by a $n$-th order form in $dx\Al$ with
$n\geq 3$,  the
problems mentioned will persist.  This discussion and the argument leading to
Eq.~\(geff)
underscore the feeling that a generic Finsler geometry is unpromising as an
arena for
matter dynamicss.  So far we have only been able to make do with the special
linear Finsler
function of Eq.~\(linearF) which is equivalent to a Riemannian metric.

But perhaps a Finsler geometry picks out a certain Riemann geometry more
complex than
$\gtab$ as special.  If so we might contemplate using that metric to construct
field
equations for matter. One would then check whether the physics is consistent
with the
Finsler geometry being the arena for matter dynamics. We shall now explore
this
program.

Our principal tool will be considerations of causality.  In order to be clear,
let us introduce the terms ``graviton'' and ``photon'' with very specific
meaning.  We have agreed that $\gab$ is the metric which the
Einstein-Hilbert action is written with.  This means that the characteristics
of the
Einstein-like equations which govern gravitational dynamics in the envisaged
theory must lie on the null surfaces of the metric $\gab$.  Short wavelength
perturbations of $\gab$ will thus propagate on these null surfaces. We call
these gravitons (no quantum conotation implied).

By photons we mean short wavelength excitations of matter fields like the
scalar field, the
Maxwell field or the Weyl neutrino field.  In GR these travel on the
null cone of the metric.  We shall adopt this wisdom as an axiom of the
present theory and take it to mean that if viewed as a point particle, a
photon follows a trajectory in the Finsler geometry with $ds^2=0$.
We have seen in Sec. II that such trajectories correspond to free particles in
the theory.
The totality of such
trajectories passing through a point in spacetime defines the
physical lightcone at that event.  Note that we do not assume that the null
surfaces of $\gab$ coincide with the physical lightcones
as would be the case in theories where the geometries are necessarily
conformally related.  Thus we do not assume that $F>0$
everywhere as is often done in studies of Finsler geometry\q{Roxburgh}.
We do assume -- and this is the content of the
causality principle --  that all physical particles travel on
trajectories with $ds^2\leq 0$.  It is still true here that nothing
travels faster than light.

Note that the point
$H=+\infty$ of the geometry is to be identified with $H=-\infty$.  This
is because the passage from one to the other corresponds to $\lineab$
passing through zero from negative to positive values.  Therefore, the
line element $ds^2$ should be continuous  as $H$ jumps from
$+\infty$ to $-\infty$, so that as $H\rarrow \pm\infty$, $F$ must either be
bounded or blow up no faster than linearly with $H$. If $F$ blows up linearly,
the coefficient of $H$ must be identical in both limits to preclude a
discontinuity in  $ds^2$.  We shall discount the possibility that $F$ can blow
up slower than $H$ because this would entail nonanalytic behavior, \eg
$F\sim H^{1/3}$.

\subhead{{\bf A.  Finsler Function} $F\,$ {\bf with no Zeros}}

Suppose $F$ is  of one sign throughout with no zeroes in the finite
$H$ axis. We take its
sign positive by convention  and refer to this case as Case A.  Now
the lightcone is delineated by coordinate increments which make
$\lineab = 0$. But in order for $ds^2$ to actually vanish for such increments,
$F$ is not allowed to blow up (even just linearly with $H$) as
$H\rarrow\pm\infty$ for in that case
the zero of $\lineab$ would be cancelled out.

Now massive  particles (not photons) follow trajectories with $ds^2 < 0$.  So
since $F > 0$, Eqs.~\(condensedfinsler) and \(H) tell us that $\lineab < 0$
and $H > 0$ for the corresponding trajectories.  Thus physical trajectories
fill the whole
positive $H$ axis, with photon trajectories lying at $H=\pm\infty$ at which
point $F$
must be bounded.

Suppose $F$ does not tend to zero as $H\rarrow +\infty$.   Clearly we must
require $F(I,+\infty,\psi)= F(I,-\infty,\psi)$  so that the line element does
not jump
between  $H=+\infty$ and $H=-\infty$. Then the graviton null surface coincides
with the
physical lightcone.  In fact, near the graviton null surface $ds^2\approx
F(I,\pm\infty,\psi)\,\lineab$, \ie the Finsler geometry induces a Riemann
geometry near
the lightcone because the conformal factor is $dx\Al$ independent there.

Following our program let us construct the matter physics
\ie Maxwell's equations, the gauge field equations,
Weyl's equation, \etc  using the effective metric
$$
\gtab\equiv F(I,\infty,\psi)\,\gab    \eqno(gplain)
$$
that has been picked out as special  by the Finsler geometry.
It is clear that short wavelength solutions of these field equations will
propagate on the lightcone as defined above simply because their
characteristics coincide with the null surface of the metric used to
build them.  Photons will thus travel on the lightcone so their
trajectories will extremize the action $S$ as required of classical
particles moving in the Finsler physical geometry.  Thus we reach a
consistent picture of photon dynamics.

But for massive particles a dichotomy appears.   These might be described by
the
Dirac equation with nonzero rest mass, or the massive Klein-Gordon equation.
In order that the weak equivalence principle be satisfied, let these field
equations be
formulated with the same metric $\gtab$ as used for the other fields.  The
classical
trajectories corresponding to a field equation may be inferred, say, from the
the
Hamilton-Jacobi equation that results from  the eikonal approximation to the
field equation.
Working out this procedure for the massive Klein-Gordon equation shows the
trajectories
to be geodesics of the effective metric $\gtab$.

Unless some very special conditions are satisfied\q{vdB}, these will not be
geodesics of the
Finsler geometry, \ie extrema of the action $S$.  An inconsistency thus
appears:
the two descriptions of particles predict different trajectories.  The
only way to bring about harmony is to require that
$F(I,H,\psi)=F(I,\infty,\psi)=A(I,\psi)$, \ie that the Finsler geometry reduce
to the
Riemann geometry defined in Eq.~\(gplain).  Of course,  this is just a special
case of the
physical geometry obtained in Eq.~\(geff).

The above remarks are not directly applicable to the subcase when the
Finsler function vanishes assymptotically as
$H\rarrow +\infty$.  For then the line
element $ds^2$ remains non-Riemannian on the graviton null surface, \ie
Eq.\(gplain) is not applicable.  We are left without a metric with which
to build
the field equations, so that the matter physics would remain ill
defined.  We conclude that, physically speaking,  this behavior of $F$ must be
excluded.

\subhead{{\bf B.  Finsler Function} $F\,$ {\bf With One Zero}}

Case A does not exhaust the possibilities.  The function
$F$ may have zeros in $H$ (for finite $H$).  Let us consider the
case, Case B, in which
$F$ has one zero, $H=h(I,\psi)$, being positive
for $H>h$ and  negative for $H<h$.
We see that this
zero corresponds to the physical lightcone.  That is, $ds^2=0$ when
$$
{(\psi,\al dx\Al)^2\over -\linemn}=h(I,\psi)\quad
\or\quad (h\gab+\psi,\al\,\psi,\be)\,dx\Al dx\Be=0 . \eqno(H=h)
$$
In obtaining the second equation  use has been made of the assumption
that $h\not= \infty$ \ie  $\linemn\not= 0$ at the zero of $F$.
Thus, the coordinate increments $dx\Al$
which are null with respect to the Riemann metric
$h\gab+\psi,\al\psi,\be$ have $ds^2=0$ with respect to the
Finsler geometry.  They make up the lightcone on which photons propagate.

Trajectories whose tangent coordinate
increments have $H<0$ with $H<h$ or
$H>h$  with $H>0$ are physical trajectories of ``massive''
particles (because they have $ds^2<0$).  Thus when $h\leq 0$  (the
physically interesting case as we shall see in Sec. IV),
the whole positive $H$ axis and that part of the negative axis left
of $h$ represent  trajectories of massive particles.  It may be seen that
the lightcone, at $H=h$,  is the boundary of these trajectories (recall that
$H=+\infty$
and $H=-\infty$ are to be identified), in accordance with intuition.

If $F$ is bounded as $H\rarrow \pm\infty$, another lightcone
appears: when $\lineab=0$, $ds^2$ vanishes also.
The simultaneous
existence of two lightcones at one event, with the second one having
physically acceptable trajectories on either side of it is unphysical.
We thus require that $F\sim H$ as $H\rarrow \pm\infty$
so that the vanishing of $\lineab$
is compensated for.
In this way  $ds^2\not=0$ on the null graviton surface.

In order to pick out a special metric from the Finsler geometry, let us expand
$F(I,H,\psi)$ in powers of $(H-h)$ about the lightcone
$H=h$.  Retaining only the first term we have
$$
ds^2=\lineab\,F'(I,H=h,\psi)\,(H-h) .
  \eqno(Hexpanded)
$$
If we now multiply in  $\lineab$ and define
$$
B(I,\psi)\equiv -F(I,h,\psi)  \qquad   \and
\qquad A(I,\psi)\equiv  h(I,\psi)B ,   \eqno(AandB)
$$
we find that the geometry in the vicinity of the lightcone is Riemannian with
the metric
$\gtab$ of Eq.~\(geff).   Since we are considering the case where $F>0$ for
$H>h$, we must
require that $B<0$ in this case.

How would things change had we assumed that $F$ is negative for $H>h$ ?
In that case the physically allowed trajectories are restricted to the range
$[0, h]$ of $H$.
Graviton trajectories (at $H=\pm\infty$) are not contiguous to this physical
range so that
gravitons travel on trajectories with $ds^2>0$.  We may thus
exclude this case by causality and  require
$$
B(I,\psi)<0 .  \eqno(Bnegative)
$$

The Finsler geometry has thus picked out a special Riemannian metric, $\gtab$,
whose
null surfaces coincide with the physical lightcones.  By analogy with the
discussion in the
last subsection, we must construct all field equations, like Maxwell's,
 with $\gtab$ to insure that short waves
travel on the physical lightcone, $\plineab=0$.  This
will then be in harmony with the classical ``photon'' trajectories derived from
the action
$S$. As before, massive particles will be predicted by the field equations to
travel on
geodesics of $\gtab$, and by the action $S$ to follow entirely different
trajectories --
extremal curves in the Finsler geometry.   Harmony can be secured only by
requiring that
the Finsler geometry reduce, for any $dx\Al$,  to a Riemann geometry with
metric $\gtab$.
Then the expansion Eq.~\(Hexpanded) is exact.  Note that in this case
the requirement that $F$ blow up linearly with $H$ as $H\rarrow\pm\infty$ is
met
automatically.

\subhead{{\bf  C. Other Cases}}

The other cases of the Finsler function $F$ may be  characterized by the
number and order of the zeroes it posesses in the variable $H$.  If
there are more than one zero, we return to the problem of multiple
lightcones, and must thus exclude this case.
Even when there is only one zero, at $H=h(I,\psi)$, we must face the
possibility that it is a zero of higher order, \ie that one or several
derivatives of $F$ vanish
at $H=h$.

Suppose $F$ has a zero of order $n>1$.  In its vicinity we
may  expand $F$ in powers of $H-h$ and retain the first term:
$$
ds^2=-D(I,\psi){[(h(I,\psi)\gab+\psi,\al \psi,\be)\,dx\Al dx\Be]^n \over (\gab
dx\Al dx\Be)^{n-1}} .  \eqno(odds)
$$
The manifold defined by the vanishing of the square brackets
in this expression is the lightcone.  We see that in its vicinity  the line
element is
not even approximately Riemannian.  Of course we may still use the metric
$h\gab+\psi,\al\psi,\be$ to construct Maxwell's equations, \etc and
these will give photons which travel on its null surface.
However, the usual problem will arise for classical massive particles:
the field equations suggest they follow geodesics of
$h\gab+\psi,\al\psi,\be$, but the action $S$ predicts they follow
geodesics of the Finsler geometry.  We cannot bring about harmony here
by having the Finsler geometry reduce to Riemann geometry because the
higher order zero means that  the Finsler geometry is never close to a Riemann
one, at
least not near the lightcone.  We must, therefore, exclude the case with a
higher order zero.

So far we have concentrated on special Riemann metrics picked out by the
Finsler geometry on
the basis of its behavior near the light cone.   There may be special Riemann
metrics which
are not    so characterized.  However, it is very doubtful that  they will
serve to construct
matter field equations whose behavior at the light cone is compatible with the
required
photon trajectories.  On this account we are restricted to the Riemann metrics
given by
Eq.~\(geff).

\subhead{{\bf IV.  DISFORMAL TRANSFORMATION AND GRAVITATIONAL
THEORY}}

We may subsume both of Cases A and B by adopting the form Eq.~\(geff)
for the physical metric together with the stipulation that $B(I,\psi)$
cannot be positive.  Then $\gtab$ in Eq.~\(geff) with  $B=0$ coincides
with the physical metric for Case A.  Thus on the basis of weak equivalence,
causality,
and consistency between the preditions of matter field equations and classical
action principle for the trajectories of classical particles,  we have shown
that Eq.~\(geff) is the most general relation between gravitational metric
$\gab$ and
physical metric $\gtab$.

One more refinement is in order.  The metric $\gtab$ depends explicitly
on $\psi$ as well as on its derivatives.  In general this would mean
that one cannot change the zero of $\psi$ without changing the metric in
a nonnegligible way.  One can conceive that it would be useful to retain
the property of translation in $\psi$ which exists in other contexts in
physics.
We may secure this by requiring that both $A$ and $B$ in the metric
depend on $\psi$ only through a common factor of the form $\exp(2\psi)$.
Any redefinition of the zero of $\psi$ then amounts to a multiplication
of the physical metric by a constant, \ie to a global  change of units
which is physically irrelevant.  Implementing this factoring
leads us to our final expression for the physical metric:
$$
\gtab\equiv e^{2\psi}[A(I)\gab+B(I)\psi,\al \psi,\be]
 \eqno(physicalmetric)
$$
with
$$
B(I)\leq 0 .  \eqno(Bcondition)
$$
We recognize Eq.~\(Bcondition) as the condition imposed by causality.
It has earlier been derived in a different way in
\ref{Kyoto}.

We call the sort of relation between $\gab$ and $\gtab$  in
Eq.~\(physicalmetric) a \i{disformal} transformation.  The term  is meant
as a contrasting one to \i{conformal} transformation which is the
special case with $B=0$ of Eq.~\(physicalmetric).  When $B=0$ the
transformation of a region of spacetime implied by $\gab\rarrow\gtab$
leaves all shapes invariant and merely stretches all spacetime
directions equally.  When $B\not= 0$ the stretch in the direction parallel
to $\psi,\al$ is by a different factor from that in the other spacetime
directions and shapes are distorted.  Maxwell's
equations, the Weyl equation for spinors, gauge field equations, \etc
will all be invariant under the transformation with $B=0$, but will not
be invariant under $\gab\rarrow\gtab$ with $B\not= 0$.  Of course,
the physical context in which we have introduced the disformal
transformation requires that all the mentioned equations be written at the
outset
with  the metric $\gtab$.

The disformal transformation was introduced in \ref{Kyoto} where
restrictions on the various functions appearing in it were summarized.
One is condition~\(Bcondition).  We reiterate some of the others.  For
this purpose let us define
$$
C(I)\equiv A(I)+I\,B(I) .    \eqno(C)
$$
The ratio $C/A$ quantifies the anisotropy of the disformal
transformation: the direction along $\psi,\al$ is stretched by a factor
$C/A$ as large as that for the other three directions.
Because we think of $\gab$ as a gravitational metric, it is clear its
signature must be $\{-,+,+,+\}$ globally (or $\{+,-,-,-\}$ in the competing
convention).  Otherwise,  the lack of global hyperbolicity will make the
setting up of the initial value problem for the metric $\gab$
impossible.  Furthermore, the physical metric $\gtab$ must also have
signature $\{-,+,+,+\}$ globally in order that it may be able to
reduce to a Minkowski
metric at every spacetime event.  By considering these conditions in a
local frame with one axis aligned  with $\psi,\al$, it is possible to
show that\q{Kyoto}
$$
A(I) > 0;\qquad\qquad  C(I) > 0 .   \eqno(A+C)
$$
Comparing Eqs.~\(AandB), \(Bcondition) and \(A+C) we see that necessarily
$h$ depends only on $I$ and $h(I)<0$.

To be a \i{bona fide} metric, the physical metric  $\gtab$ must be
invertible: there must be an inverse metric $\gtAB$ at every spacetime point.
If it exists it must be of the form
$$
\gtAB=e^{-2\psi}A(I)^{-1}[\gAB-L^2B(I)C(I)^{-1}g^{\alpha\mu}g^{\beta\nu}
\psi,\m \psi,\n] .     \eqno(inverse)
$$
The conditions \(A+C) guarantee that this inverse is well defined
everywhere.

One more condition can be obtained if it is agreed that there should be a
one-to-one correspondence between gravitational metric and physical
metric.  Suppose one contracts Eq.~\(inverse) with $\psi,\be$.  The
result is
$$
\gAB\psi,\be=C(I)\,e^{2\psi}\gtAB\psi,\be .  \eqno(eigen)
$$
A  further contraction with $\psi,\al$  gives
$$
J=e^{-2\psi}IC(I)^{-1} ,  \eqno(J)
$$
where by $J$ we mean the analog of $I$ written with metric $\gtAB$:
$$
J\equiv  L^2\,\gtAB\psi,\al \psi,\be .   \eqno(Jdef)
$$

In principle it is possible to solve Eq.~\(J) for $I$.
If we solve Eq.~\(physicalmetric) for $\gab$ and eliminate $I$ everywhere,
we get
$$
\gab=A[I(J\,e^{2\psi})]^{-1}\{e^{-2\psi}\gtab-L^2B[I(J\,e^{2\psi})]
\psi,\al
\psi,\be\} .     \eqno(gfromgtwiddle)
$$
It can be seen that if $I(J\,e^{2\psi})$ were to be multiply valued, there
would be
several gravitational metrics for each physical metric, which would be
unphysical.  The way to
avoid this is to require that $J$ be a monotonic function of $I$ for
fixed $\psi$.  We shall require
$$
{d\over dI}\Big[{I\over C(I)}\Big]>0 .      \eqno(condition)
$$
We have required $I/C$ to be increasing because the opposite assumption
runs counter to the situation in many known theories (see below).

\subhead{{\bf V. CONCLUSIONS AND QUESTIONS}}

We thus conclude that subject to conditions \(Bcondition), \(A+C) and
\(condition),
Eq.~\(physicalmetric) is the most general relation between    and
gravitational metric $\gab$ which respects the weak equivalence principle,
ordinary
notions of causality, and which is insensitive to a change of zero for the
auxiliary
scalar field $\psi$.

Finsler geometry served mostly a negative role in our argument.  Although it is
more
general than Riemann geometry, we found it rather unpromising for building
dynamical
equations for matter.  It did point us to certain Riemannian geometries as
candidates for
the physical geometry.  Had one proceeded entirely within the framework of
Riemann
geometry, one might not have thought of a relation like Eq.~\(physicalmetric)
between
gravitational and physical geometries. Thus the line of thought developed here
opens up
for discussion  a broader class of physical geometries than those conformal to
the
gravitational geometry.

A conformal transformation between metrics can be interpreted as a change
of local units of length\q{Dicke, Dirac}: the ratio between
gravitational and material units varies from event to event.  By analogy
we may interpret
the disformal transformation as a change of local units of length
for which the units for intervals along the gradient of $\psi$ are
different than those for  intervals orthogonal to it.  Conformal
transformations have traditionally been the source of insight into field
theory.  It  appears likely that disformal transformations will
help to supplement those insights.  At the most immediate level, they
provide a method for constructing novel gravitational theories based on
pairs of disformally related geometries.
As far as we are aware, such  theories
have been considered only once before\q{Kyoto},  and the motivation there was
to relate
the standard interpretation of the data from gravitational lenses to
the modified gravity
resolution of the missing mass puzzle\q{Milgrom}.  The present work thus
provides a
theoretical backbone for  those studies.

Thus far we have only described a framework; concrete theories will
arise when dynamics are specified for $\psi$.  GR is the trivial case;
it
corresponds to the requirement that $\psi=\const$. By an appropriate
choice of units it can be arranged that $\exp(2\psi)\,A(0)=1$, so that,
as expected,
GR equates physical and gravitational geometry. The value of $B$ is
plainly irrelevant and may be set to zero by fiat.
Brans-Dicke theory (in Dicke's form)\q{Dicke}, Dirac's theory\q{Dirac}, and
the variable mass theory\q{VMT} all prescribe nontrivial dynamics for $\psi$,
but choose (in the appropriate units) $A(I)=1$
and $B(I)=0$.  One can further conceive of theories with $A(I) \not=\const$
and $B(I)<0$
in which gravitons travel  slower than photons, a feature which might be
subject to direct
experimental test.  One such theory has been studied in detail in \ref{Kyoto}.

   If the true
gravitational theory is of the conventional type ($A=1$ and $B=0$), the
question
arises what symmetry or selection principle forces $A$ and $B$ into these
trivial values
when they could be functions of the invariant $I$ ?  Conversely, if nature has
taken
advantage of the wider possibilities for $A$ and $B$, in what ways are the
intuitions
about gravity that have been molded by conventional theories to be modified ?
Further
work will take up these and other questions.

 \subhead{{\bf ACKNOWLEDGMENTS}}

I would like to thank Jim Hartle and Gary Horowitz for hospitality at Santa
Barbara.
\endtitlepage
\References
\oneandathirdspace
\parskip=2pt plus 3pt
\refis{Nordstrom}\obscure{G. Nordstr{\o}m}{Ann. Phys.}{42}{533}{13}.

\refis{Cartan}\book{E. Cartan}{Les Espaces de Finsler}
{Herman, Paris}{34}.

\refis{Roxburgh}\grg{I. W. Roxburgh}{23}{1071}{91}.

\refis{Roxburgh2}\grg{I. W. Roxburgh}{24}{4191}{92}.

\refis{Dirac}\obscure{P. A. M. Dirac}{Proc. Roy. Soc.
London}{A333}{403}{73}.

\refis{VMT}\prd{J. Bekenstein}{15}{1458}{77}; \prd{J. D. Bekenstein and
A. Meisels}{18}{1313}{80}.

\refis{Dicke}\pr{C. Brans and R. H. Dicke}{124}{925}{61};
\pr{R. H. Dicke}{125}{2163}{62}.

\refis{strings}\npb{C. G. Callan, R. C. Myers and M. J. Perry}
{311}{673}{88}.

\refis{Schild}\inbooked{A. Schild}{Relativity Theory and
Astrophysics}{J. Ehlers}{American Mathematical Society, Providence,
Rhode Island}{67}.

\refis{Kyoto}\inbooked{J. D. Bekenstein}{The Sixth Marcel Grossmann Meeting
on General
Relativity}{H. Sato}{World Publishing, Singapore}{92}.

\refis{vdB}\grg{R. K. Tavakol and N. van den Bergh}{18}{849}{86}.

\refis{Milgrom}\apj{M. Milgrom}{270}{365}{83} ; \apj{J. D.
Bekenstein and M. Milgrom}{286}{7}{84}; \aar{R. H. Sanders}{2}{1}{90}.

\endreferences

\endit\end